\documentclass[aps,10pt,nofootinbib, preprintnumbers, showpacs]{revtex4}

\usepackage{amssymb, amsfonts, amsmath, bm}

\begin{document}

\title{{\Large
Construction of test Maxwell fields with scale symmetry
}}
\author{{\large Takahisa Igata}}
\email{igata@rikkyo.ac.jp}
\affiliation{Department of Physics, Rikkyo University, Toshima, Tokyo 175-8501, Japan}
\date{\today}
\preprint{RUP-18-13}
\pacs{04.20.-q, 03.50.De, 11.30.-j}

\begin{abstract}
Geometrical symmetry in a spacetime can generate 
test solutions to the Maxwell equation. 
We demonstrate that the source-free Maxwell equation is 
satisfied by any generator of spacetime 
self-similarity---a proper homothetic vector---identified 
with a vector potential of the Maxwell theory. 
The test fields obtained in this way 
share the scale symmetry of the background. 
\end{abstract}

\maketitle

\textit{Introduction.}---A test field 
on a background spacetime can be a good approximation 
when its contribution to the stress--energy--momentum tensor 
is negligibly small. 
In particular, test Maxwell fields play a large role 
to model various astrophysical phenomena, 
such as Blandford--Znajek process~\cite{Blandford:1977ds} 
(see Refs.~\cite{Toma:2014kva, Kinoshita:2017mio} for recent progress),  
black hole Meissner effect~\cite{King:1975tt, Bicak:1985, Takamori:2010mb}, etc. 
In most cases, 
assuming that test fields share symmetry of the background spacetime, 
we find test solutions 
by solving the Maxwell equation.

However, we also have a simple and powerful tool 
for obtaining test Maxwell solutions 
from spacetime symmetry~\cite{Wald:1974np}. 
The symmetry refers to an isometry of the spacetime metric
and is generated by a Killing vector. 
Because of geometrical properties relevant to the symmetry, 
an arbitrary Killing vector 
in a vacuum spacetime (i.e., Ricci flat) can
be a test vector potential to the source-free Maxwell equation. 
Such electromagnetic fields obtained in this way 
share the symmetry generated by the Killing vector. 
Some of them can be models of physical interest.  
Therefore the solutions have been widely used, e.g., 
in the construction of electromagnetic field solutions 
on the Kerr black hole background~\cite{Koide:1999bj, 
Igata:2010ny, Igata:2012js, Shiose:2014bqa, Shaymatov:2014dla}.

Furthermore, we can relate 
a Killing--Yano 2-form~\cite{Yano:1952} or  
a conformal Killing--Yano 
2-form~\cite{Tachibana:1968, Kashiwada:1969, Tachibana:1969} to 
a test field-strength in the Maxwell theory. 
A closed conformal Killing--Yano 2-form 
solves the Maxwell equation sourced by 
the electric current density that 
is proportional to a Killing vector derived from it. 
This is known as the Killing--Maxwell system~\cite{Carter:1987id} 
and can be found, e.g., in the 4D Kerr spacetime 
because a Killing--Yano 2-form exists~\cite{Penrose:1973um, Floyd:1973}.%
\footnote{The dual of a Killing--Yano 2-form is 
a closed conformal Killing--Yano 2-form~\cite{Yasui:2011pr, Frolov:2017kze}. }
On the other hand, 
the combination of a Killing--Yano 2-form and the Riemann tensor 
can provide Maxwell's field-strength that solves the source-free Maxwell equation in a vacuum spacetime~\cite{Hughston:1990}. 
This relation was generalized to 
the combination of 
a conformal Killing--Yano 2-form and a spin-2 field~\cite{Jezierski:2002mn}. 
Recently, in any Einstein space with a closed conformal Killing--Yano 2-form, 
a new procedure for obtaining a solution to the source-free Maxwell equations 
was proposed~\cite{Frolov:2017bdq}. 
These many examples mean that 
geometrical symmetry in a spacetime is 
closely related to the solutions of the Maxwell equations in various ways.

In this short note, 
we focus on the relation of test Maxwell fields
and scale symmetry of a spacetime, 
which is generated by a proper homothetic vector. 
The spacetime admitting the symmetry
is known as a self-similar spacetime.  
The self-similar spacetimes are found at the critical point of 
the critical phenomena in gravitational collapse~\cite{Choptuik:1992jv, Koike:1995jm} 
and appears in spherically symmetric cosmological models~\cite{Carr:1998at} and  
in spatially homogeneous cosmologies~\cite{Wainwright:1997}. 
Recently, classical particle and string mechanics in a self-similar spacetime 
were studied in the relation of self-similarity to 
conserved quantities and self-similar configurations~\cite{Igata:2016uvp, Igata:2018dxl}. 
The purpose of this short note is to 
develop the procedure for obtaining 
test Maxwell fields by means of 
a proper homothetic vector.

This this is organized as follows. 
In the following section, we derive an identity between 
the second derivative of a homothetic vector and the Riemann tensor. 
Then we show that the solution construction of the vacuum test Maxwell equation 
by means of the Killing vector carries over the homothetic vector, 
i.e., 
an arbitrary homothetic vector satisfies the source-free Maxwell equation
in a vacuum self-similar spacetime. 
Furthermore, 
we demonstrate to construct 
a self-similar test Maxwell solution 
to the source-free Maxwell equation
in the 4D Kasner spacetime. 
In the last section, 
we summarize our results. 
Throughout this note we use the abstract index notation~\cite{Wald:1984rg}.

\textit{Main results.}---Let $ (M, g_{ab})$ 
be a $D$-dimensional self-similar spacetime. 
Then the metric~$g_{ab}$ admits 
a proper homothetic vector $\xi^a$, which satisfies
\begin{align}
\label{eq:HVeq}
\pounds_\xi g_{ab}\equiv \nabla_a \xi_b+\nabla_b \xi_a =2\:\!g_{ab},
\end{align}
where $\pounds_\xi$ is the Lie derivative with respect to $\xi^a$ 
and $\nabla_a$ is the Levi-Civita covariant derivative associated with $g_{ab}$. 
In what follows, 
we consider a relation between the second derivative of $\xi^a$ and 
the Riemann tensor~$R_{abc}{}^d$. 
By definition, we can relate the second derivative of $\xi^a$ to $R_{abc}{}^d$ as
\begin{align}
\nabla_a \nabla_b \:\!\xi_c -\nabla_b \nabla_a\:\! \xi_c=R_{abc}{}^d \xi_d.  
\end{align}
Using Eq.~\eqref{eq:HVeq} to the second term on this left-hand side, we have
\begin{align}
\nabla_a \nabla_b \:\!\xi_c+\nabla_b \nabla_c\:\! \xi_a=R_{abc}{}^d \xi_d,
\end{align}
where we have used the compatibility condition $\nabla_a g_{bc}=0$. 
From cyclic permutations of the indices of this equation, we have 
\begin{align}
\label{eq:KRrelation}
\nabla_a \nabla_b\:\! \xi_c
=\frac{1}{2}\left(R_{abc}{}^d+R_{cab}{}^d-R_{bca}{}^d\right)\xi_d
=-R_{bca}{}^d \xi_d,
\end{align}
where we have used the identity of the Riemann tensor $R_{[\:\!abc\:\!]}{}^d=0$ in the last equality. Contracting Eq.~\eqref{eq:KRrelation} over $a$ and $b$, 
we obtain
\begin{align}
\label{eq:HVRicci}
\nabla_a \nabla^a \xi^b=-R^b{}_a \xi^a. 
\end{align}
where $R_{ab}=R_{acb}{}^c$ is the Ricci tensor. 
Note that 
these relations between a homothetic vector and spacetime curvatures 
is the same as those between a Killing vector and spacetime curvatures.

Now, we introduce a closed 2-form field from $\xi_a$ as 
\begin{align}
\label{eq:F}
F_{ab}=2\:\!\nabla_{[\:\!a} \xi_{b\:\!]}
=2\left(
\nabla_a \xi_b-g_{ab}
\right),
\end{align}
where we have used Eq.~\eqref{eq:HVeq} in the last equality.  
Then the relation \eqref{eq:HVRicci} is written in terms of $F_{ab}$ as
\begin{align}
\nabla_a F^{ab} =-2\:\!R^{b}{}_a \:\!\xi^a,
\end{align}
where $\nabla_a g_{bc}=0$ was used. 
Hence, the homothetic vector $\xi_a$ 
satisfies the source-free Maxwell equation if 
either of the following conditions is satisfied: 
(i) $\xi^a$ is an eigen vector of the Ricci tensor with 
a zero eigen value or
(ii) the spacetime is vacuum, i.e., the Ricci tensor vanishes. 
Indeed, there exist self-similar vacuum spacetimes 
with whole-cylinder symmetry~\cite{Harada:2008rx} 
and with spatial homogeneity 
in cosmologies (see, e.g., Ref.~\cite{Wainwright:1997}). 
If $\xi_a$ is closed (i.e., a pure gauge), $F_{ab}$ vanishes.  
The gauge of the vector potential $\xi_a$ is partially fixed 
because Eq.~\eqref{eq:HVeq} leads to 
\begin{align}
\nabla_a \xi^a=D. 
\end{align}
Though the right-hand side is not zero, 
this may be a kind of Lorenz condition 
in the sense that the equation of motion reduces source-free wave equation
in the Ricci flat case.

The symmetry of the Maxwell field is 
induced from the scale symmetry of the metric on the background. 
The Lie derivative of the homothetic vector $\xi_a$ 
with respect to $\xi^a$ yields 
\begin{align}
\pounds_\xi \xi_a=2\:\!\xi_a. 
\end{align}
Hence the exterior derivative of this equation leads to 
\begin{align}
\pounds_\xi F_{ab}=2\:\!F_{ab},
\end{align}
where the commutability of the exterior derivative and the Lie derivative was used. 
This means that $F_{ab}$ is a self-similar 2-form with weight $2$, i.e., 
the Maxwell field has self-similarity. 
As is apparent from the derivation, the self-similar weight is related to 
that of the metric fixed in Eq.~\eqref{eq:HVeq}.

The total electric charge $Q$ of the electromagnetic field~\eqref{eq:F} 
in 3D volume $\Sigma$ is given by
\begin{align}
Q=\frac{1}{4\pi }\int_{\partial \Sigma} {}^*\!F_{ab}
=\frac{1}{4\pi} \int_{\partial \Sigma} \nabla^a\xi^b\epsilon_{abcd}
=\frac{1}{2\pi}\int_{\Sigma} R^{d}{}_{e}\xi^{e}\epsilon_{dabc},
\end{align}
where we have used Eq.~\eqref{eq:HVRicci} in the last equality. 
This is the Komar integral for the homothetic vector $\xi^a$.

As an example of the above procedure, 
we obtain a test Maxwell field associated with a homothetic vector 
in a spatially homogeneous self-similar vacuum spacetime. 
All spatially homogeneous solutions to the vacuum Einstein equations 
that admit a 4D homothetic group $H_4$ acting simply transitively on spacetime 
are classified in Ref.~\cite{Hsu:1986ri}. 
Except for the Minkowski spacetime, 
the 4D Kasner spacetime~\cite{Kasner:1925} is 
the simplest background, 
which means that $H_4$ contains the 3D isometry subgroup $G_3=\mathbb{R}^3$. 
The Kasner spacetime appears in 
each epoch of Belinskii--Khalatnikov--Lifshitz oscillation near 
cosmological singularity~\cite{Belinsky:1970ew} or
the past asymptotic behavior of some Bianchi universes. 
Note that the analytical continuation of the 
Kasner spacetime yields the Levi-Civita spacetime~\cite{Kastor:2015wda}, 
which is static and cylindrically symmetric vacuum spacetime~\cite{Levi-Civita:1919}.

We find test Maxwell fields with scale symmetry in the 4D Kasner spacetime. 
The spacetime metric is given by
\begin{align}
\mathrm{d}s^2
=-\mathrm{d}t^2
+t^{2\:\!p_1}\:\!\mathrm{d}x^2
+t^{2\:\!p_2}\:\!\mathrm{d}y^2
+t^{2\:\!p_3}\:\!\mathrm{d}z^3,
\end{align}
where the parameters $p_1$, $p_2$, and $p_3$ are constrained by the equations%
\begin{align}
\label{eq:parametercondition1}
p_1+p_2+p_3=1, 
\quad
p_1^2+p_2^2+p_3^2=1. 
\end{align}
Since the Kasner spacetime is self-similar~\cite{McIntosh:1991},
we find a proper homothetic vector
\begin{align}
\label{eq:HVKasner}
\xi^a
=t\left(\partial/\partial t\right)^a
+\left(1-p_1\right) x\left(\partial/\partial x\right)^a
+\left(1-p_2\right) y \left(\partial/ \partial y\right)^a
+\left(1-p_3\right) z\left(\partial /\partial z\right)^a.
\end{align}
When we set any one of $p_i$ equal to unity, 
we find the other parameters being zero from Eqs.~\eqref{eq:parametercondition1} 
(e.g., $p_1=1$, $p_2=0$, and $p_3=0$). 
Then the metric in this case reduces the 2D Rindler spacetime crossed with $\mathbb{R}^2$, i.e., 
the 4D Minkowski spacetime after the analytic extension, and $\xi_a$
reduces closed. 
We obtain $F_{ab}$ in Eq.~\eqref{eq:F} by using 
Eq.~\eqref{eq:HVKasner}
\begin{align}
F_{ab}=4  
p_1\left(1-p_1\right)\:\!t^{2\:\!p_1-1}\:\!x\:\!(\mathrm{d}t)_{[\:\!a}\:\!(\mathrm{d} x)_{b\:\!]}
+4p_2\left(1-p_2\right)\:\!t^{2\:\!p_2-1}\:\!y\:\!(\mathrm{d}t)_{[\:\!a}\:\!(\mathrm{d}y)_{b\:\!]}
+4p_3\left(1-p_3\right)\:\!t^{2\:\!p_3-1} \:\!z\:\!(\mathrm{d}t)_{[\:\!a}\:\!\mathrm{d}z_{b\:\!]}.
\end{align}
It follows immediately from this form that 
$F_{ab}$ is degenerate, i.e., $F_{ab}{}^*\!F^{ab}=0$.  
Though this is a pure electric solution, 
we can obtain more physically interesting solutions, 
e.g., a pure magnetic solution by a duality rotation of $F_{ab}$. 
The electric charge $Q$ vanishes because of the 
Ricci flatness of the Kasner spacetime. 
The electromagnetic field is dynamical due to 
the cosmological background. 
The role of magnetic fields were widely
investigated in Bianchi cosmologies (see, Ref.~\cite{LeBlanc:1997} for a review). 

\textit{Summary.}---We have revealed a relation between 
spacetime scale symmetry and test Maxwell fields. 
Identifying a homothetic vector in a self-similar spacetime
with a vector potential in the Maxwell theory, 
we can obtain a source-free test 
Maxwell field solution. 
This procedure is the same as the well-known for 
constructing a test Maxwell solution using a Killing vector. 
Whether or not the solutions obtained in this way 
is physically interesting is a different issue 
from the relation of spacetime symmetry and test Maxwell fields. 
It will be interesting to find a solution of physical interest 
in this framework for future work. 

\begin{acknowledgments}
This work was supported by the MEXT-Supported Program for the Strategic Research Foundation at Private Universities, 2014-2017 (S1411024). 
\end{acknowledgments}


\begin{thebibliography}{99} 
%
\bibitem{Blandford:1977ds} 
  R.~D.~Blandford and R.~L.~Znajek,
  Electromagnetic extractions of energy from Kerr black holes,
  Mon.\ Not.\ Roy.\ Astron.\ Soc.\  {\bf 179}, 433 (1977).  

%
\bibitem{Toma:2014kva} 
  K.~Toma and F.~Takahara,
  Electromotive force in the Blandford--Znajek process,  
  Mon.\ Not.\ Roy.\ Astron.\ Soc.\  {\bf 442}, 
  2855 (2014)
  [arXiv:1405.7437 [astro-ph.HE]].

%
\bibitem{Kinoshita:2017mio} 
  S.~Kinoshita and T.~Igata,
  The essence of the Blandford--Znajek process,
  PTEP {\bf 2018}, 
  033E02 (2018)
  [arXiv:1710.09152 [gr-qc]].
  
%
\bibitem{King:1975tt} 
  A.~R.~King, J.~P.~Lasota, and W.~Kundt,
  Black Holes and Magnetic Fields,
  Phys.\ Rev.\ D {\bf 12}, 3037 (1975).
  
\bibitem{Bicak:1985}
  J.~Bicak and V.~Janis,
  Magnetic fluxes across black holes,
  Mon.\ Not.\ Roy.\ Astron.\ Soc.\  {\bf 212}, 899 (1985).
  
  
%
\bibitem{Takamori:2010mb} 
  Y.~Takamori, K.~i.~Nakao, H.~Ishihara, M.~Kimura, and C.~M.~Yoo,
  Perturbative Analysis of a Stationary Magnetosphere in an Extreme Black Hole Spacetime : On the Meissner-like Effect of an Extreme Black Hole,
  Mon.\ Not.\ Roy.\ Astron.\ Soc.\  {\bf 412}, 2417 (2011)
  [arXiv:1010.4104 [gr-qc]].
  
%
\bibitem{Wald:1974np} 
  R.~M.~Wald,
  Black hole in a uniform magnetic field,
  Phys.\ Rev.\ D {\bf 10}, 1680 (1974).

%
\bibitem{Koide:1999bj} 
  S.~Koide, D.~L.~Meier, K.~Shibata, and T.~Kudoh,
  General relativistic simulations of jet formation in a rapidly rotating black hole magnetosphere,
  Astrophys.\ J.\  {\bf 536}, 668 (2000)
  [astro-ph/9907435].

%
\bibitem{Igata:2010ny} 
  T.~Igata, T.~Koike, and H.~Ishihara,
  Constants of Motion for Constrained Hamiltonian Systems: A Particle around a Charged Rotating Black Hole,
  Phys.\ Rev.\ D {\bf 83}, 065027 (2011)
  [arXiv:1005.1815 [gr-qc]].
  
%
\bibitem{Igata:2012js} 
  T.~Igata, T.~Harada, and M.~Kimura,
  Effect of a Weak Electromagnetic Field on Particle Acceleration by a Rotating Black Hole,
  Phys.\ Rev.\ D {\bf 85}, 104028 (2012)
  [arXiv:1202.4859 [gr-qc]].

\bibitem{Shiose:2014bqa} 
  R.~Shiose, M.~Kimura, and T.~Chiba,
  Motion of Charged Particles around a Weakly Magnetized Rotating Black Hole,
  Phys.\ Rev.\ D {\bf 90}, 
  124016 (2014)
  [arXiv:1409.3310 [gr-qc]].
  
\bibitem{Shaymatov:2014dla} 
  S.~Shaymatov, M.~Patil, B.~Ahmedov, and P.~S.~Joshi,
  Destroying a near-extremal Kerr black hole with a charged particle: Can a test magnetic field serve as a cosmic censor?,
  Phys.\ Rev.\ D {\bf 91}, 
  064025 (2015)
  [arXiv:1409.3018 [gr-qc]].    
  
\bibitem{Yano:1952}
  K.~Yano,
  Some remarks on tensor fields and curvature, 
  Ann.\ Math.\ {\bf 55}, 328 (1952). 

\bibitem{Tachibana:1968}
  S.~Tachibana,
  On Killing tensors in a Riemannian space,
  Tohoku Math.\ J.\ {\bf 20}, 257 (1968). 
  
\bibitem{Kashiwada:1969}
  T.~Kashiwada, 
  On conformal Killing tensor,
  Nat.\ Sci.\ Rep.\ Ochanomizu Univ.\ {\bf 19}, 67 (1968).
  
\bibitem{Tachibana:1969}
  S.~Tachibana and T.~Kashiwada, 
  On the integrability of Killing--Yano's equation, 
  J.\ Math.\ Soc.\ Japan {\bf 21}, 259 (1969). 
  
%
\bibitem{Carter:1987id} 
  B.~Carter,
  Separability of the Killing--Maxwell System Underlying the Generalized Angular Momentum Constant in the Kerr--Newman Black Hole Metrics,
  J.\ Math.\ Phys.\  {\bf 28}, 1535 (1987).

%
\bibitem{Penrose:1973um} 
  R.~Penrose,
  Naked singularities,
  Annals N.\ Y.\ Acad.\ Sci.\  {\bf 224}, 125 (1973).
  
\bibitem{Floyd:1973}
  R.~Floyd, 
  The dynamics of Kerr fields, 
  PhD Thesis University of London (1973). 


%
\bibitem{Yasui:2011pr} 
  Y.~Yasui and T.~Houri,
  Hidden Symmetry and Exact Solutions in Einstein Gravity,  
  Prog.\ Theor.\ Phys.\ Suppl.\  {\bf 189}, 126 (2011)
  [arXiv:1104.0852 [hep-th]].
  
%
\bibitem{Frolov:2017kze} 
  V.~Frolov, P.~Krtous, and D.~Kubiznak; 
  Black holes, hidden symmetries, and complete integrability; 
  Living Rev.\ Rel.\  {\bf 20}, 
  6 (2017)
  [arXiv:1705.05482 [gr-qc]].
  
  
\bibitem{Hughston:1990}
  L.~P.~Hughston and K.~P.~Tod, 
  An Introduction to General Relativity, 
  Cambridge University Press (1990). 

%
\bibitem{Jezierski:2002mn} 
  J.~Jezierski,
  CYK tensors, Maxwell field and conserved quantities for the spin-2 field,
  Class.\ Quant.\ Grav.\  {\bf 19}, 4405 (2002)
  [gr-qc/0211039].  
  
%
\bibitem{Frolov:2017bdq} 
  V.~P.~Frolov, P.~Krtous, and D.~Kubiznak,
  Weakly charged generalized Kerr--NUT--(A)dS spacetimes,
  Phys.\ Lett.\ B {\bf 771}, 254 (2017)
  [arXiv:1705.00943 [gr-qc]].

%
\bibitem{Koike:1995jm} 
  T.~Koike, T.~Hara, and S.~Adachi,
  Critical behavior in gravitational collapse of radiation fluid: A Renormalization group (linear perturbation) analysis,
  Phys.\ Rev.\ Lett.\  {\bf 74}, 5170 (1995)
  [gr-qc/9503007].

%
\bibitem{Choptuik:1992jv} 
  M.~W.~Choptuik,
  Universality and scaling in gravitational collapse of a massless scalar field,
  Phys.\ Rev.\ Lett.\  {\bf 70}, 9 (1993).

%
\bibitem{Carr:1998at} 
  B.~J.~Carr and A.~A.~Coley,
  Self-similarity in general relativity,
  Class.\ Quant.\ Grav.\  {\bf 16}, R31 (1999)
  [gr-qc/9806048].

\bibitem{Wainwright:1997}
  J.~Wainwright and G.~F.~R.~Ellis,
  Dynamical Systems in Cosmology,
  Cambridge University Press (1997). 

%
\bibitem{Igata:2018dxl} 
  T.~Igata,
  Scale invariance and constants of motion, 
  arXiv:1804.03369 [hep-th].
  
%
\bibitem{Igata:2016uvp} 
  T.~Igata, T.~Houri, and T.~Harada,
  Self-similar motion of a Nambu-Goto string,
  Phys.\ Rev.\ D {\bf 94}, 
  064029 (2016)
  [arXiv:1608.03698 [gr-qc]].
  
\bibitem{Wald:1984rg} 
  R.~M.~Wald,
  General Relativity,
  University of Chicago Press (1984). 

%
\bibitem{Harada:2008rx} 
  T.~Harada, K.~i.~Nakao, and B.~C.~Nolan,
  Einstein--Rosen waves and the self-similarity hypothesis in cylindrical symmetry,
  Phys.\ Rev.\ D {\bf 80}, 024025 (2009), 
  Erratum: [Phys.\ Rev.\ D {\bf 80}, 109903 (2009)]
  [arXiv:0812.3462 [gr-qc]].

%
\bibitem{Hsu:1986ri} 
  L.~Hsu and J.~Wainwright,
  Self similar spatially homogeneous cosmologies: Orthogonal perfect fluid and vacuum solutions,
  Class.\ Quant.\ Grav.\  {\bf 3}, 1105 (1986).

\bibitem{Kasner:1925}
  E.~Kasner, 
  Solutions of the Einstein Equations Involving Functions of Only One Variable, 
  Trans.\ Am.\ Math.\ Soc.\ {\bf 27}, 155 (1925). 

%
\bibitem{Belinsky:1970ew} 
  V.~A.~Belinskii, I.~M.~Khalatnikov, and E.~M.~Lifshitz,
  Oscillatory approach
   to a singular point in the relativistic cosmology,
  Adv.\ Phys.\  {\bf 19}, 525 (1970).

\bibitem{Levi-Civita:1919}
  T.~Levi-Civit\`a, 
  Einsteinian $ds^2$ in Newtonian fields. IX: The analog of the logarithmic potential, 
  Gen.\ Relativ.\ Gravit.\ {\bf 43}, 2321 (2011) [republication of Rend.\ Acc.\ Lincei {\bf 28}, 101 (1919)]. 

%
\bibitem{McIntosh:1991}
  C.~B.~G.~McIntosh and J.~D.~Steele, 
  All Vacuum Bianchi I Metrics with a Homothety, 
  Classical Quantum Gravity {\bf 8}, 1173 (1991). 

  
\bibitem{LeBlanc:1997}
  V.~G.~LeBlanc, 
  Asymptotic states of magnetic Bianchi I cosmologies,
  Classical Quantum Gravity {\bf 14}, 2281 (1997).

%
\bibitem{Kastor:2015wda} 
  D.~Kastor and J.~Traschen,
  Melvin Magnetic Fluxtube/Cosmology Correspondence,
  Class.\ Quant.\ Grav.\  {\bf 32}, 
  235027 (2015)
  [arXiv:1507.05534 [hep-th]].
\end{thebibliography}
\end{document}